\documentclass{revtex4}
\begin{document}
\title{\textbf{On the possibility to use non-orthogonal orbitals  for
Density Matrix Renormalization Group calculations in Quantum Chemistry}}
\author{A. O. Mitrushenkov}
\address{Department of Theoretical Physics, Institute of Physics, St. Petersburg University, 198904 St. Petersburg, Russia}
\author{Guido Fano}
\address{Dipartimento di Fisica, Universit\`{a} di Bologna, Via Irnerio 46, 40126 Bologna, Italy}
\author{Roberto Linguerri and Paolo Palmieri}
\address{Dipartimento di Chimica Fisica ed Inorganica, Universit\`{a} di Bologna,
Viale Risorgimento 4, 40136 Bologna, Italy}
\date{February 10 2003}

\begin{abstract}
The generalization of Density Matrix Renormalization Group (DMRG)
approach as implemented in quantum chemistry, to the case of
non-orthogonal orbitals is carefully analyzed. This generalization
is attractive from the physical point of view since it allows
a better localization of the orbitals. The possible
implementation difficulties and drawbacks are estimated. General
formulae for hamiltonian matrix elements useful in DMRG calculations
are given.
\end{abstract}

\maketitle
\parskip=0.5cm
\parindent=0cm


\section{Introduction}

The DMRG method, first introduced by White \cite{White1,White2,WhiteDMRG93}
in solid state physics, has been very successful in calculations of
 low dimensional
quantum lattice systems, and
 has become recently an interesting tool  in quantum chemistry
 \cite{PariserFano,Bendaz,WhiteDMRGqc1,WhiteDMRGqc2,FanoLing,Gordon,
Hess}.  The method is attractive because of the good scaling properties
of the calculation time with the size of the system \cite{WhiteDMRGqc1}.
\newline \indent In the present work we first list the main steps of DMRG algorithm
as implemented in quantum chemistry. All previous implementations
were restricted to the orthonormalized molecular orbitals (MOs).
While rendering all algorithms much simpler, the standard MOs may
be not quite appropriate from the physical point of view,
especially if we try to follow the original DMRG ideas that assume
the separation of the system into well defined physical blocks.
From this point of view the use of bond structures, e.g. valence
bond orbitals \cite{Thorst} can be much more attractive as it allows for
clear interpretation of resulting wave functions. Even if it is
not clear that this can improve the results numerically, we
consider below the theoretical aspects of the use of
non-orthogonal orbitals as applied to the DMRG algorithms.
\newline \indent The article is organized as follows: first we describe essential
steps of the classical, orthogonal, DMRG algorithm. In section
\ref{sect2} we describe the generalization to the case of
non-orthogonal orbitals. In appendices we give the full list of
relevant formulae.

\section{Classical, orthogonal,  DMRG algorithm\label{sect1}}

We do not describe here the details of the DMRG approach, it can be
found in many reviews \cite{WhiteDMRG93,PariserFano,Gordon}
. We also do not describe the
details of implementation of DMRG in quantum chemistry, see \cite{WhiteDMRGqc1,
FanoLing}.
Rather we concern only with practical, computational, aspect of
the algorithm.
\newline The main points of DMRG procedure are:
\begin{itemize}
\item The parametrization of the wave function \item List of
`primitive' matrix elements stored \item Adding orbitals\item
Calculation of Hamiltonian matrix elements\item Diagonalization of
Hamiltonian matrix\item Expanded space states construction
\end{itemize}

\subsection{General form of the wave function in DMRG}

In DMRG we divide the whole system into $A$ and $U$ subsystems
(two subsets of the full set of MOs in quantum chemistry). The
wave function is represented as
\begin{equation}\label{eq1}
    \Psi=\sum_{IJ}C_{IJ}|A_I\otimes U_J>
\end{equation}
where $|A_I>$ describes some state of subsystem $A$ and similar
for $U$. The tensor product $A_I\otimes U_J$ is automatically antysymmetrized
due to the presence of the creation operators (see \cite{FanoLing}
  for details).
 The conservation rules imply for nonzero $C_{IJ}$
\begin{equation}\label{eq2}
    n_\alpha(A_I)+n_\alpha(U_J)=n_\alpha
\end{equation}
where $n_\alpha$ is fixed by the total number of electrons and
total spin projection, the same being true for $\beta$. If there
present additional spatial symmetries, the corresponding
conservation rules must be also satisfied.
\newline One important drawback of the above expression is that the wave
function given by Eq.(\ref{eq1}) is not necessarily the
eigenfunction of total spin operator $\hat S^2$. We actually never
considered the application of corresponding projection operator to
get the spin-pure DMRG wavefunction, maybe it is not really as
difficult as the $\hat S^2$ is the combination of
creation/annihilation operators whose matrix elements are in
general available in DMRG (see below).

\subsection{State pair types and matrix elements to be
stored}\label{ss2}

To find the wave function given by Eq.(\ref{eq1}) we must
diagonalize Hamiltonian matrix in the space span by $|A_I\otimes
U_J>$ state vectors. To build this matrix  all we need are some
matrix elements for pure $A$ and $U$ blocks that we call
`primitive' matrix elements. The set of these matrix elements
depends on the relations between number of $\alpha$ and $\beta$
electrons. In the following when we consider particular matrix
element
\begin{equation}\label{eq3}
    <IJ|\hat H|KL>
\end{equation}
indices $I$ and $K$ refer to $A$ subsystem while $J$ and $L$ - to
$U$ subsystem.
\newline Let's denote with the symbol $\textbf{a}^{\dagger}$ a creation operator
referring to block \emph{A} and  with $\textbf{u}^{\dagger}$ a creation operator
referring to block \emph{U}.
The electronic part of quantum chemistry Hamiltonian can be written as
\begin{equation}\label{aeq1}
    \hat H=\hat H_A + \hat H_U + \hat H_{AU} ,
\end{equation}
where $\hat H_A$,  $\hat H_U$ contain only $\textbf{a}$, $\textbf{a}^{\dagger}$ or
$\textbf{u}$, $\textbf{u}^{\dagger}$ operators and the interaction term is
given by
\begin{equation}\label{aeq2}
    \hat H_{AU} = \sum_{p q x y} [(xy|pq)-(xq|py)][\textbf{u}^{\dagger}_{x\uparrow}
\textbf{u}_{y\uparrow}\textbf{a}^{\dagger}_{p\uparrow}\textbf{a}_{q\uparrow}+
\textbf{u}^{\dagger}_{x\downarrow}
\textbf{u}_{y\downarrow}\textbf{a}^{\dagger}_{p\downarrow}\textbf{a}_{q\downarrow}]+
(xy|pq)[\textbf{u}^{\dagger}_{x\uparrow}
\textbf{u}_{y\uparrow}\textbf{a}^{\dagger}_{p\downarrow}\textbf{a}_{q\downarrow}+
\textbf{u}^{\dagger}_{x\downarrow}
\textbf{u}_{y\downarrow}\textbf{a}^{\dagger}_{p\uparrow}\textbf{a}_{q\uparrow}],
\end{equation}
where $p,q$ indices refer to orbitals of $A$ block, $x,y$ to $U$.
\newline Then, given $I$ and $K$, the Table \ref{tab1} lists the 7 cases
(others can be obtained using $<IJ|\hat H|KL>=<KL|\hat H|IJ>$) for
which the matrix elements Eq.(\ref{eq3}) are nonzero. Please note
that the $JL$-case is inverse relative to that of $IK$, i.e. the
case of $LJ$ is the same as $IK$.

\begin{table}
\caption{7 cases for $IK$-relations}\label{tab1}
\begin{tabular}{|c|c|c|}
  \hline
  Case & $\alpha$ & $\beta$ \\
  \hline
  1 & $n_I^\alpha=n_K^\alpha$ & $n_I^\beta=n_K^\beta$ \\
  2 & $n_I^\alpha=n_K^\alpha+1$ & $n_I^\beta=n_K^\beta-1$ \\
  3 & $n_I^\alpha=n_K^\alpha-1$ & $n_I^\beta=n_K^\beta$ \\
  4 & $n_I^\alpha=n_K^\alpha$ & $n_I^\beta=n_K^\beta-1$ \\
  5 & $n_I^\alpha=n_K^\alpha-2$ & $n_I^\beta=n_K^\beta$ \\
  6 & $n_I^\alpha=n_K^\alpha$ & $n_I^\beta=n_K^\beta-2$ \\
  7 & $n_I^\alpha=n_K^\alpha-1$ & $n_I^\beta=n_K^\beta-1$ \\
  \hline
\end{tabular}
\end{table}

With this definition, the `primitive' matrix elements needed to
build the full Hamiltonian matrix, depend on the `case' given
above. The complete list of these matrix elements is given in
Appendix \ref{app1}, together with resulting matrix elements of
the total Hamiltonian (see below). Here we just note that all ME's
have the typical form of
\begin{equation}\label{eq4}
    <I|\hat a_{p\alpha}\dots|K>
\end{equation}
with one to three creation/annihilation operators, or the linear
combination of such ME's.

\subsection{Calculation of Hamiltonian matrix elements}

After we have defined the stored `primitive' matrix elements in
the previous subsection, the matrix elements of Hamiltonian matrix
look like
\begin{equation}\label{eq5}
    <IJ|\hat H|KL>=\sum_i f_i(IK)g_i(JL)
\end{equation}
where type and range of index $i$ depend on the $IK$-type. The
full set of matrix elements is given in Appendix \ref{app1}, where
the bi-orthogonal orbitals are assumed, see next sections. For the
standard orthogonal case the dual orbitals coincide with original
orbitals so the tildes in the formulas of Appendix \ref{app1} can
be simply omitted.

\subsection{Adding orbitals}

Adding of orbital(s) is the most important and basic step of the
whole DMRG procedure. Given the set of $A$ states (the same is
valid for $U$, so we do not speak of it further), we add few
orbitals to $A$ and build the new bigger set of states including
all possible occupations of added orbital pattern. Indices $Ii$,
$Jj$, etc. refer to the new enlarged subsystems. The subsequent
step is the recalculation of the `primitive' matrix elements in
this new  set of states. The only practical difficulty
here is that the number of cases to treat grows considerably.
\newline In our early implementations we recalculated the `primitive'
matrix elements after adding orbitals and stored them on disk.
This led to significant reductions in performance. Therefore
later, to save disk space and improve performance, we changed our
algorithms as not to calculate and to store the extended set of
`primitive' matrix elements but rather to build the Hamiltonian
matrix in the extended space explicitly using the non-extended
`primitive' matrix elements and some additional molecular
integrals. Also, the highly optimized
$BLAS$\cite{blas1,blas2,blas3,blas4,blas5} routines are used as
much as possible to make the code faster. Note that this part of
code can be easily parallelized. The full list of relevant
formulas is given in Appendix \ref{app2}. The tildes can be safely
omitted again there.

\subsection{Hamiltonian diagonalization in the orthogonal case}

Once the Hamiltonian matrix elements in $|A_I\otimes U_J>$ basis
are available, the next task is to diagonalize this Hamiltonian
matrix. We can not apply the explicit techniques like Jacobi
diagonalization, because of quite big, tens or hundreds of
thousands states, dimension of typical problem. Therefore we are forced
to use the direct technique, which  is based on
the ability to calculate the so called $\sigma$-vector, i.e.
action of Hamiltonian
\begin{equation}\label{eq6}
    {\bf y}=\hat H{\bf x}
\end{equation}
given vector ${\bf x}$ without explicit storage of Hamiltonian
matrix $H_{\alpha\beta}$.
In the usual formulation with orthogonal orbitals,  Davidson method is the most efficient.
As well known, it consists of an iterative procedure that starts from some reasonable guess and
then expands the set of trial vectors up to say 30-50 that provide
good approximation to exact eigenvector.
The procedure can be seen as a minimization of the Rayleigh quotient
\begin{equation}\label{eq8}
    E=\frac{<{\bf x}|\hat H|{\bf x}>}{<{\bf x}|{\bf x}>},
\end{equation}
in which one applies the Newton method and replace
 $ D = \hbox{diag H} $
in the Hessian matrix.

\subsection{New states construction}

After diagonalizing Hamiltonian matrix in the extended set of
vectors, the next and the last task for the whole DMRG procedure
is the construction of reduced subset of states describing $A$ and
$U$ in the extended orbital set. This is done by finding the best
possible approximation to the calculated eigenstate of Hamiltonian
matrix when the number of states retained is fixed to a number ($M$)
smaller than the
total dimension spanned by $|Ii\otimes Jj>$; this approximation
is obtained by minimizing
\begin{equation}\label{eq10}
    |\Psi-\Psi_{\hbox{trial}}|^2
\end{equation}

In practice it is done by performing the Singular Value
Decomposition (SVD) of the matrix $C_{IJ}$ (see Eq.(2.1))
which gives the components of the wave function
in the $ A_I \otimes U_J$ basis, and then selecting
the largest singular values and the
corresponding states.
\newline After we know the expansion of selected $M$ states in terms of
$|Ii>$ for $A$ and $|Jj>$ for $U$, we must recalculate and store
the `primitive' matrix elements for these new states. Again it is
done `on the fly', expanding the matrix elements to newly added
orbitals and contracting at the same time with coefficients
produced by SVD procedure. For reference purposes in Appendix
\ref{app3} we list all expanded matrix elements. These formulas
are closely related to that of Appendix \ref{app2}.

\section{Non-orthogonal orbitals\label{sect2}}

In this section we consider how the algorithms of the previous
section can be extended to the case of non-orthogonal molecular
orbitals.
\newline We start by describing general techniques to work with
non-orthogonal orbitals. In principle there exist a general
L\"owdin formula that expresses the matrix elements of one- and
two-electron operators between two determinants build with
non-orthogonal orbitals as the sum of appropriate matrix
determinants. It is this necessity to evaluate many matrix
determinants that renders this approach completely unusable in
practice. The practical way to work with non-orthogonal orbitals
consists in introducing the so called bi-orthogonal (dual)
orbitals \cite{Douall,Thorstein}.

\subsection{General remarks on the eigenvalue problem with non-orthogonal
bases}

First of all, we cannot use the variational approach for the
simple reason that we are unable to evaluate directly the diagonal
energy expression which we demonstrate in the following
section\footnote{This also means that even if we can still get the
energy following the procedure described below, we will not be
able to evaluate any expectation values. Or at least it requires
some additional work that we do not consider here.}.
\newline In order to compute the variational expression (Eq.(2.9)) for the
energy, we must be able to express the vector $ |\Psi> $ in
the biorthogonal orbitals;  if we were able to find this expansion,
then the computation follows more or less the standard root with
insignificant changes since the one-electron and two-electron
integrals loose their transformation properties.  However in practice
to find and use such expansion is an almost impossible task. In fact
when the orbitals of $A$ and $U$ do overlap, a given state of the block $A$
would become a linear combination (of nearly full CI size) of different
$A$  and $U$  states.  Therefore the whole strategy of DMRG becomes
unapplicable. Still we will show that
some modern non-variational approach can be used with
 more or less success.
\newline Let us introduce some notations that will make clearer the
following discussion. Given a set $\{\phi_i\}$ of non orthogonal
orbitals, the {\it dual} or {\it biorthogonal} basis $\tilde\phi_i$
verifies the condition:
\begin{equation}
    < \tilde \phi_i | \phi_j > = \delta_{ij}
\end{equation}
Let $S$ denote the overlap matrix $<\phi_i |\phi_j>$ ; given two
one-particle states $|a> = \sum_i a^i | \phi_i >$, $|b>= \sum_i b^i |\phi_i>$,
we can define two different scalar products: the ``physical'' scalar product
\begin{equation}
     < a | b > = \sum_{ij} a^i b^j S_{ij}
\end{equation}
which will be denoted by Dirac brackets,
and a simple ``numerical'' scalar product defined by:
\begin{equation}
      ( a | b ) = \sum_i  a^i b^i
\end{equation}
which will be denoted by round brackets. Furthermore for any state
$|a> = \sum_i a^i | \phi_i>$
we define the {\it dual} state $ |\tilde a> = \sum_i a^i |\tilde \phi_i >$;
therefore we have the identity:
\begin{equation}
      < \tilde a | b > = ( a | b )
\end{equation}
     The concept of dual state can be generalized: a {\it dual} many
particle state is the state built with the same CI (or DMRG) expansion
coefficients but with $\{\phi_i \}$ orbitals replaced by dual ones
\footnote{If the orbital overlap matrix is essentially
different from the unit matrix, the `dual' vector is completely
different from the state itself.If though the overlap matrix is
close to unit, the `dual' vector is close to original vector.}.\par
      A nice property of the biorthogonal formalism is that the
``dual'' annihilation operators defined by
\begin{equation}
      \tilde a_{i \sigma} = \sum_k a_{i \sigma} ( S^{-1})_{ki}
\end{equation}
obey the usual anticommutation relations \cite{Douall}:
\begin{equation}
      \{ a^+_{i \sigma}  \tilde a_{j \sigma} \}_+ = \delta_{\sigma \tau}
\delta_{ij}
\end{equation}
       The one-particle and two-particle integrals are computed with
the $\phi_i$ wave functions on the right and the dual $\tilde\phi_i$
on the left.  For instance:
\begin{equation}
(\tilde i j | \tilde k l ) = \int \tilde \phi_i (1) \tilde \phi_k (2)
{ 1 \over r_{12} } \phi_j (1) \phi_l (2) dr_1 dr_2
\end{equation}
In non-variational approach instead of looking for the minimum of
energy in the trial subspace we project the energy eigenvalue
equation onto the `dual' subspace.   To find the current
approximation to energy and vector given the trial vector set we
solve the following non-symmetric eigenvalue problem
\begin{equation}\label{eq12}
    H_{ij}C_j=EC_j
\end{equation}
where $H_{ij}=<\tilde{\bf x}_i|\hat H|{\bf x}_j>=({\bf x}_i|\hat
H|{\bf x}_j)$; because of biorthogonality
$<\tilde {\bf x_i} | {\bf x_j} > = ({\bf x}_i|{\bf x_j})=\delta_{ij}$.
If expansion set becomes reasonably full for the description of
target state, this non-variation approach must produce reasonably
good result, which is also demonstrated by some experience with
non-orthogonal Full CI calculations.\par
   In principle, if we want to set up a variational calculations with
non orthogonal orbitals, we need the knowledge of the transformation
operator $\hat T $ from normal to dual vectors :
\begin{equation}\label{eq13}
    {\bf x}=\hat T{\bf{\tilde x}}
\end{equation}
Then the variational energy becomes
\begin{equation}\label{eq14}
    E=\frac{<{\bf x}|\hat H|{\bf x}>}{<{\bf x}|{\bf x}>}=
    \frac{<{\bf{\tilde x}}|\hat T^{T}\hat H|{\bf x}>}{<{\bf{\tilde x}}|\hat T^{T}|{\bf
    x}>}=
    \frac{({\bf x}|\hat T^{T}\hat H|{\bf x})}{({\bf x}|\hat T^{T}|{\bf x})}
\end{equation}
An iterative minimization procedure for this energy expression
 can be easily written
down;
however, it is almost impossible to follow the variational approach.
  Even the simple square  $ <{\bf x} | {\bf x}>$ of
the physical norm of a many particle state $ |{\bf x}> $ which appears in the
denominator of Eq.(3.10), is a huge expansion of determinants, since
the scalar product of two simple configurations gives rise to a determinant;
neglecting the spin indexes, and denoting by $|0>$ the vacuum, we have:
\begin{equation}
<0|a_{i_n}....a_{i_1} a^+_{j_1}...a^+_{j_n} |0> =
\hbox{det} ( < \phi_{i_h} | \phi_{j_k} >) = \hbox{det} ( S_{i_h j_k} )
\end{equation}
As we shall see in the next paragraph, the same kind of difficulties
exist for DMRG calculations.\par
Once the non symmetric Hamiltonian matrix $H_{ij}$ is constructed, we are
left with the problem of using Davidson method or a similar direct meyhod.
In paragraph 2.5 we have seen that the Davidson method can be considered
as a minimization procedure of the expectation value (2.9). This procedure
looses any meaning for a non symmetric matrix, since the minimum of (2.9)
does not coincides with the lowest eigenvalue.  However the Davidson
 method can be seen from a different angle, and can
be used for solving the eigenvalue problem of non
symmetric matrices.  Of course the orthonormalization can be performed
only with the ``numerical'' scalar product (3.4).\par
    As well known for non symmetric matrices, there are other good methods that
can be tried. The simplest one is perhaps the two-sided or non-symmetric
Lanczos method that treats on an equal footing the Krylov sequences
$\{ {\bf x}, H {\bf x}, H^2 {\bf x}... \}$ and
$\{ {\bf x}, H^T {\bf x}, (H^T)^2 {\bf x}... \}$ generated by the matrix
$H$ and its transposed $H^T$. Let us denote by $V$ and $W$ the two
supspaces generated by the two Krylov sequences. It is easy to find
two biorthogonal bases $\{v_j\}$ and $\{w_i\}$ in $V$ and $W$, in such
a way that the matrix $ (w_i | H v_j)$ is tridiagonal
\cite{Vorst}. \par
     More complicated methods, like the two-sided Jacobi-Davidson method,
which makes use of biorthogonal basis like the two-sided Lanczos, can
also be considered \cite{Vorst} . The use of orthogonal bases
is generally more numerically stable, while two-sided methods (and biorthogonal
basis) may provide faster convergence.

\subsection{Variational calculations with non-orthogonal
orbitals in DMRG?}

Here we show that variational calculations are completely
impractical for non-orthogonal orbitals. We consider the overlap
matrix just. The Hamiltonian matrix would be even much more
complicated.
So the problem is to evaluate
\begin{equation}\label{eq16}
    <IJ|KL>
\end{equation}

Actually the problem is not just to evaluate Eq(\ref{eq16}) but do it
efficiently. For that we must factorize this expression to
something containing only pair quantities, as it is anyway
impossible to store and use $M^4$ quantities, corresponding to
Eq(\ref{eq16}). This means that we would like to have something
like
\begin{equation}\label{eq16a}
    <IJ|KL>=x_1(IJ)x_2(KL)+x_3(IK)x_4(JL)+x_5(IL)x_6(JK)
\end{equation}

It is rather easy to show  that such  factorization is
impossible in general case. We show that by considering the most
simple example when all states $I,J,K,L$ are determinants, and the
number of electrons in $I$ and $K$ states is the same (we consider
say $\alpha$ electrons, $\beta$ being similar). Note that while
for orthogonal orbitals the block structure is automatic, i.e.
number of electrons in $I$ and $K$ (and also in $J$ and $L$) MUST
be the same to get non-zero overlap, it is NOT the case for
non-orthogonal orbitals. Then the overlap, according to L\"owdin
formula is given by the determinant
\begin{equation}\label{eq16b}
    <IJ|KL>=\hbox{det} \left(%
\begin{array}{cc}
  S_{IK} & S_{IL} \\
  S_{JK} & S_{JL} \\
\end{array}%
\right)
\end{equation}
where $S$-matrices are buildee by the corresponding orbital overlaps.
Assuming the general case when matrix $S_{IK}$ is non-degenerate and
using the Gauss formula we have
\begin{equation}\label{eq16c}
    <IJ|KL>=\hbox{det}(S_{IK})\hbox{det}(S_{JL}-S_{JK}S^{-1}_{IK}S_{IL})
\end{equation}
which has a complicated mixed structure and cannot be
represented in the factorized form Eq(\ref{eq16a}).

\subsection{Non-variational calculations}

As we have seen that the only way is to perform non-variational
calculations, the rest is more or less straightforward. To
evaluate the matrix elements of Hamiltonian we use the same
expressions as for orthogonal orbitals with minor modifications
due to loss of integral symmetry. The full list of matrix elements
to build the Hamiltonian matrix if given in Appendices \ref{app1}
and \ref{app2}.
Having Hamiltonian matrix elements (or better corresponding
$\sigma$-routines), we solve non-symmetric eigenvalue problem as
described before.
The only remaining point is the reduction of states. We cannot
calculate the reduced density for the same reason for which we cannot
calculate theoverlap matrices. Actually, the concept of reduced
density matrix is not clear in the case of non-orthogonal $A$ and
$U$. The matrix
\begin{equation}
    \rho_{II'} = \sum_{J=1}^M  C_{IJ} C_{I'J}
\end{equation}
is not the reduced density matrix of one block and does not verifies
$ \hbox{Tr} [\rho] = 1 $ .\par
However we can follow a simpler approach: we can abandon the idea of
approximating the wave function using the physical norm  $<x|x>$
and we can use the following norm induced by the simple expression
(3.4):
\begin{equation}\label{eq17}
    |x|^2_{DMRG}\equiv<{\bf{\tilde x}}|{\bf x}>
  = < {\bf{\tilde x}}|\hat T {\bf{\tilde x}}>
  =({\bf x}|{\bf x})=\sum_i|x_i|^2
\end{equation}
Of course the norm defined by Eq.(\ref{eq17}) is a true norm, so
it is positive, it is zero only when ${\bf x}$ is zero ecc. Then
we can find the `best' approximation using this norm, which as
before leads to SVD decomposition of the wave function coefficient
matrix $C_{IJ}$. Again when the orbital overlap matrix resembles unit, this
should give results close to those obtained using the true physical norm.
However the minimization with the ``numerical'' norm can give good results
in more general cases.\par
Let us consider in more detail the practical side of the
calculations. Let us define
  $ OP(\sim\leftrightarrow no\sim)$ the operator that interchanges normal
dual orbitals in all the one- and two-electron integrals.  Clearly we have:
\footnote{We always work with real orbital,
so no complex conjugation is needed}
\begin{equation}\label{eqmn1}
  OP(\sim\leftrightarrow no\sim)
    [<IJ|\hat H|KL>] = <KL|\hat H|IJ>
\end{equation}

This means that all `primitive' matrix elements not containing
integrals are the same for $IK$ and $KI$. Therefore in practice we
can keep the same number of cases (7) and just calculate two
copies (one normal and one dual) of those primitive MEs that
contain integrals, i.e. $H_{IK}$ for case 1, and $f_x$ for cases 3
and 4. Then we can use Eq.(\ref{eqmn1}) to calculate inverse
cases. The attention we must pay is that when we have $LJ$ instead
of $JL$ referring to primitive MEs, we must take the dual copies.
Considering that in practice the most of storage is taken by
$f_{pq}$ primitive matrix elements of the case 1, which remains
unchanged, the bi-orthogonal case has essentially the same
memory-and-disk requirements as the standard orthogonal procedure.

\appendix
\section{Non-extended DMRG matrix elements}\label{app1}

We start by explaining some notations we use in the below tables.
The phases as as follows: the $A$ is created first, and $U$ next.
Within the state the $\alpha$- (spin-up) electrons are created
first, and $\beta$- (spin-down) next. So the $|IJ>$ state where
$I$ refers to $A$ and $J$ to $U$, can be written as
\begin{equation}
    |IJ>=|I(\uparrow)I(\downarrow)J(\uparrow)J(\downarrow)>
\end{equation}

As to letters, $I$ and $K$ will refer to $A$ block and $J$ and $L$
- to $U$. Orbital labels $p,q,r,s,\dots$ will refer to $A$
orbitals while $x,y,z,\dots$ - to $U$.
The states are ordered as follows in the program: $I<K$ if:
\begin{itemize}
\item $n_I^\beta<n_K^\beta$ \item $n_I^\beta=n_K^\beta$ and
$n_I^\alpha<n_K^\alpha$ \item $n_I^\beta=n_K^\beta$ and
$n_I^\alpha=n_K^\alpha$ and $symm(I)<symm(K)$
\end{itemize}

Here we list the matrix elements needed to be kept and the
corresponding contributions to the matrix elements of total
Hamiltonian. As noted before we consider only the case $I\le K$.
The opposite case can be obtained from the modified `Hermitian'
property (3.19).
For the specific case $1$, when $J>L$ in the below expressions, we can use
the property:
\begin{equation}\label{eqa1e2}
    f^{\alpha,\beta}_{xy}(JL)=
  OP(\sim\leftrightarrow no\sim)
[f^{\alpha,\beta}_{yx}(LJ)]
\end{equation}

Table \ref{tab2} gives all matrix elements.

\begin{table}[h]
\caption{Hamiltonian matrix elements for DMRG}\label{tab2}
\begin{tabular}{|c|c|c|c|c|}
  \hline
  Case ($I\le K$) & MEs to keep & $<IJ|\hat H|KL>$ & $J\le L?$ & Comments \\
  \hline
  1 & $<I|\hat H_A|K>$ & $\delta_{JL}<I|\hat H_A|K>+$ &  YN & \\
    & $f^\alpha_{pq}(IK)=<I|\hat a_{p\alpha}^+\hat
    a_{q\alpha}|K>$ & $\delta_{IK}<J|\hat H_U|L>+$& & \\
    & $f^\beta_{pq}(IK)$ - similar $\alpha\leftrightarrow\beta$ & $\sum_{xy}g^\alpha_{xy}(IK)f^\alpha_{xy}(JL)+$& & \\
    & $g^\alpha_{xy}(IK)=\sum_{pq}f^\alpha_{pq}(IK)\left[(\tilde pq|\tilde xy)-(\tilde py|\tilde xq)\right]$&
    $\sum_{xy}g^\beta_{xy}(IK)f^\beta_{xy}(JL)$& & Not kept\\
    & $+\sum_{pq}f^\beta_{pq}(IK)(\tilde pq|\tilde xy)$& & & (evaluated\\
    & $g^\beta_{xy}(IK)$ - similar $\alpha\leftrightarrow\beta$ & & & when needed)\\
  \hline
  2 & $f_{pq}(IK)=<I|\hat a_{p\alpha}^+\hat a_{q\beta}|K>$ & $\sum_{xy}g_{xy}(IK)f_{xy}(LJ)$ & N & \\
    & $g_{xy}(IK)=-\sum_{pq}f_{pq}(IK)(\tilde px|\tilde yq)$ & &  & Not kept \\
  \hline
  3 & $f_p(IK)=<I|\hat a_{p\alpha}|K>$ & $(-1)^{n_K^\alpha+n_K^\beta+1}\left\{\right.$ & N & \\
    & $f_x(IK)=\sum_{pqr}<I|\hat a_{q\beta}^+\hat a_{p\beta}\hat a_{r\alpha}|K>(\tilde xr|\tilde qp)+$ &
    $\sum_pf_p(IK)f_p(LJ)+$ & & \\
    & $\sum_{q;p>r}<I|\hat a_{q\alpha}^+\hat a_{p\alpha}\hat
    a_{r\alpha}|K>\left[(\tilde xr|\tilde qp)-(\tilde xp|\tilde qr)\right]$ & $\left.\sum_xg_x(IK)f_x(LJ)\right\}$&  & \\
    & $g_x(IK)=f_x(IK)+\sum_pf_p(IK)h_{\tilde px}$ & & & Not kept \\
  \hline
  4 & like 3 with $\alpha\leftrightarrow\beta$&  & N & \\
  \hline
  5 & $f_{pq}(IK)=<I|\hat a_{p\alpha}\hat a_{q\alpha}|K>$ & $\sum_{x>y}g_{xy}(IK)f_{xy}(LJ)$ & N & $p>q$ only\\
    & $g_{xy}(IK)=\sum_{q<p}f_{pq}(IK)\left[(\tilde xp|\tilde yq)-(\tilde yp|\tilde xq)\right]$ & &  & $x>y$ only; Not kept \\
  \hline
  6 & like 5 with $\alpha\leftrightarrow\beta$ &  & N & \\
  \hline
  7 & $f_{pq}(IK)=<I|\hat a_{p\alpha}\hat a_{q\beta}|K>$ & as for Case 2 & N & \\
    & $g_{xy}(IK)=\sum_{pq}f_{pq}(IK)(\tilde xp|\tilde yq)$ & &  & Not kept \\
  \hline
\end{tabular}
\end{table}

\section{Extended DMRG Hamiltonian matrix elements}\label{app2}

In this Appendix we give the matrix elements of Hamiltonian in the
extended basis, i.e. when we have added the orbitals to all four
states $I,J,K,L$. The states over those added orbitals are simple
determinants, which we denote as patterns $i,j,k,l$. The number of
cases thus grows enormously if we count for all possible
occupations combinations for states $I,J,K,L$ and patterns
$i,j,k,l$. The fact that states over $i,j,k,l$ are complete, i.e.
are simple determinants, allows us to reformulate the problem by
shifting everything to $A$ block, i.e. we call the `combined'
pattern $i+j$ as new $i$ and $k+l$ as new $k$. The only thing that
remains after we have diagonalized Hamiltonian matrix is to cast
this combined representation back to original split one. This is
just the phase and indexing problem that is done very efficiently.
Thus we have now orbitals added only to $A$ block, as $i$ and $k$
patterns and thus the number of cases becomes reasonable. The
phase convention is as follows:
\begin{equation}
    |IiJ>=|I(\uparrow)I(\downarrow)i(\uparrow)i(\downarrow)J(\uparrow)J(\downarrow)>
\end{equation}

The orbital labels $a,b,c,\dots$ will refer to orbitals of
patterns $i$ and $k$. In below expressions the sums over those
indices $a,b,c,\dots$ is assumed when not given explicitly. Table
\ref{tab3} lists all the matrix elements. For the explanation of
quantities $f$ and $g$ see Table \ref{tab2}.
The `new' case $IiKk$ means the relations between number of
electrons in $Ii$ and $Kk$ blocks, in the same way as before. The
overscore on the case index mean inverted case, i.e. $Ii>Kk$.

\begin{table}[h]
\caption{Extended Hamiltonian matrix elements for
DMRG}\label{tab3}
\begin{tabular}{|c|c|c|c|c|}
  \hline
  Old case ($IK$) & New case ($IiKk$) & $(Ii)\le(Kk)$? & $J\le L$?
  & $<IiJ|\hat H|KkL>$ \\
  \hline\hline
  1 & 1 & YN & YN & $\delta_{ik}<IJ|\hat H|KL>+\delta_{IK}\delta_{JL}<i|\hat H|k>+$\\
    &&&& $\delta_{i_\alpha
    k_\alpha}\delta_{IK}\sum_{ab}g_{ab}^\beta(JL)f_{ab}^\beta(i_\beta
    k_\beta)+$ \\
    &&&&
    $\delta_{i_\alpha k_\alpha}\delta_{JL}\sum_{ab}g_{ab}^\beta(IK)f_{ab}^\beta(i_\beta
    k_\beta)+$\\
    &&&& $\delta_{i_\beta k_\beta}\delta_{IK}\sum_{ab}g_{ab}^\alpha(JL)
    f_{ab}^\alpha(i_\alpha k_\alpha)+$ \\
    &&&& $\delta_{i_\beta k_\beta}\delta_{JL}\sum_{ab}g_{ab}^\alpha(IK)f_{ab}^\alpha(i_\alpha
    k_\alpha)$ \\
  \hline
  1 & 2 & Y  &  N & $\delta_{IK}(-1)^{n_k^\alpha}\sum_{ab}<i_\alpha|\hat a^+_{a\alpha}|k_\alpha>
                    <i_\beta|\hat a_{b\beta}|k_\beta>g_{ab}(LJ)$\\ \hline
  1 & 3 & Y  &  N & $(-1)^{n_k^\alpha+n_k^\beta+1}\left\{\delta_{IK}\left[
     \delta_{i_\beta k_\beta}\left(\sum_ag_a(LJ)<i_\alpha|\hat a_{a\alpha}|k_\alpha>+\right.\right.\right.$\\
  &&&& $\left.\sum_xf_x(LJ)\sum_{a,b>c}<i_\alpha|\hat
  a_{a\alpha}^+\hat a_{b\alpha}\hat
  a_{c\alpha}|k_\alpha>[({\tilde a}b|{\tilde x}c)-({\tilde a}c|{\tilde x}b)]\right)+$\\
  &&&& $\left.\sum_xf_x(LJ)\sum_{abc}<i_\alpha|\hat
  a_{c\alpha}|k_\alpha><i_\beta|\hat a_{a\beta}^+\hat
  a_{b\beta}|k_\beta>({\tilde a}b|{\tilde x}c)\right]+$\\
  &&&& $\left.\delta_{i_\beta
  k_\beta}\sum_{cx}g_{xc}^\alpha(IK)f_x(LJ)<i_\alpha|\hat
  a_{c\alpha}|k_\alpha>\right\}$\\
  \hline
  1 & 4 & Y  &  N & $(-1)^{n_k^\beta+1}\left\{\delta_{IK}\left[
     \delta_{i_\alpha k_\alpha}\left(\sum_ag_a(LJ)<i_\beta|\hat a_{a\beta}|k_\beta>+\right.\right.\right.$\\
  &&&& $\left.\sum_xf_x(LJ)\sum_{a,b>c}<i_\beta|\hat
  a_{a\beta}^+\hat a_{b\beta}\hat
  a_{c\beta}|k_\beta>[({\tilde a}b|{\tilde x}c)-({\tilde a}c|{\tilde x}b)]\right)+$\\
  &&&& $\left.\sum_xf_x(LJ)\sum_{abc}<i_\beta|\hat
  a_{c\beta}|k_\beta><i_\alpha|\hat a_{a\alpha}^+\hat
  a_{b\alpha}|k_\alpha>({\tilde a}b|{\tilde x}c)\right]+$\\
  &&&& $\left.\delta_{i_\alpha
  k_\alpha}\sum_{cx}g_{xc}^\beta(IK)f_x(LJ)<i_\beta|\hat
  a_{c\beta}|k_\beta>\right\}$\\
  \hline
  1 & 5 & Y  &  N & $\delta_{IK}\delta_{i_\beta k_\beta}
                    \sum_{a>b}<i_\alpha|\hat a_{a\alpha}\hat a_{b\alpha}|k_\alpha>g_{ab}(LJ)$\\ \hline
  1 & 6 & Y  &  N & $\delta_{IK}\delta_{i_\alpha k_\alpha}
                    \sum_{a>b}<i_\beta|\hat a_{a\beta}\hat a_{b\beta}|k_\beta>g_{ab}(LJ)$\\ \hline
  1 & 7 & Y  &  N & $\delta_{IK}(-1)^{n_k^\alpha}\sum_{ab}<i_\alpha|\hat a_{a\alpha}|k_\alpha>
                    <i_\beta|\hat a_{b\beta}|k_\beta>g_{ab}(LJ)$\\ \hline\hline
  2 & 1 & YN & YN & $\delta_{JL}(-1)^{n_k^\alpha+1}\sum_{ab}<i_\alpha|\hat a_{a\alpha}|k_\alpha>
                    <i_\beta|\hat a^+_{b\beta}|k_\beta>g_{ab}(IK)$\\ \hline
  2 & 2 & Y  &  N & $\delta_{ik}\sum_{xy}g_{xy}(IK)f_{xy}(LJ)$ \\ \hline
  2 & \=3 & N  &  Y & $\delta_{i_\alpha
  k_\alpha}(-1)^{n_k^\beta}\sum_{c}<i_\beta|\hat
  a^+_{c\beta}|k_\beta>\sum_xf_x(JL)g_{xc}(IK)$\\ \hline
  2 & 4 & Y  &  N & $\delta_{i_\beta
  k_\beta}(-1)^{n_k^\alpha+n_k^\beta+1}\sum_{c}<i_\alpha|\hat
  a_{c\alpha}|k_\alpha>\sum_xf_x(LJ)g_{cx}(IK)$\\ \hline\hline
  3 & 1 & YN & YN & $(-1)^{n_K^\alpha+n_K^\beta+1}\left\{\delta_{JL}\left[
     \delta_{i_\beta k_\beta}\left(\sum_ag_a(IK)<i_\alpha|\hat a^+_{a\alpha}|k_\alpha>+\right.\right.\right.$\\
  &&&& $\left.\sum_pf_p(IK)\sum_{a,b>c}<i_\alpha|\hat a^+_{c\alpha}\hat a^+_{b\alpha}\hat
  a_{a\alpha}|k_\alpha>[({\tilde b}a|{\tilde c}p)-({\tilde c}a|{\tilde b}p)]\right)+$\\
  &&&& $\left.\sum_pf_p(IK)\sum_{abc}<i_\alpha|\hat
  a^+_{c\alpha}|k_\alpha><i_\beta|\hat a_{b\beta}^+\hat
  a_{a\beta}|k_\beta>({\tilde b}a|{\tilde c}p)\right]+$\\
  &&&& $\left.\delta_{i_\beta
  k_\beta}\sum_{cp}g_{cp}^\alpha(JL)f_p(IK)<i_\alpha|\hat
  a^+_{c\alpha}|k_\alpha>\right\}$\\
  \hline
  3 & \=2 & N  &  Y & $\delta_{i_\alpha
  k_\alpha}(-1)^{n_k^\alpha+n_K^\alpha+n_K^\beta+1}<i_\beta|\hat
  a^+_{c\beta}|k_\beta>\sum_{cp}f_p(IK)g_{pc}(JL)$\\ \hline
  3 & 3 & Y & N &
  $(-1)^{n_k^\beta+n_k^\alpha+n_K^\alpha+n_K^\beta+1}\left\{\right.$\\
  &&&& $\delta_{ik}\left(\sum_pf_p(IK)f_p(LJ)+\sum_xg_x(IK)f_x(LJ)\right)+$\\
  &&&& $\delta_{i_\alpha k_\alpha}\sum_{ab}<i_\beta|\hat
  a_{a\beta}^+\hat
  a_{b\beta}|k_\beta>\sum_{px}f_p(IK)f_x(LJ)({\tilde x}p|{\tilde a}b)+$\\
  &&&& $\left.\delta_{i_\beta k_\beta}\sum_{ab}<i_\alpha|\hat
  a_{a\alpha}^+\hat
  a_{b\alpha}|k_\alpha>\sum_{px}f_p(IK)f_x(LJ)[({\tilde x}p|{\tilde a}b)-({\tilde a}p|{\tilde x}b)]\right\}$\\\hline
  3 & \=3 & N & Y &
  $(-1)^{n_k^\beta+n_k^\alpha+n_K^\alpha+n_K^\beta}\delta_{i_\beta k_\beta}$\\
  &&&&$\sum_{a>b}<i_\alpha|\hat a^+_{b\alpha}\hat
  a^+_{a\alpha}|k_\alpha>\sum_{px}f_p(IK)f_x(JL)[({\tilde a}p|{\tilde b}x)-({\tilde
b}p|{\tilde a}x)]$\\\hline
  3 & 4 & Y & N &
  $(-1)^{n_k^\beta+n_K^\alpha+n_K^\beta}\sum_{ab}<i_\alpha|\hat a^+_{a\alpha}|k_\alpha>$\\
  &&&&$<i_\beta|\hat a_{b\beta}|k_\beta>\sum_{px}f_p(IK)f_x(LJ)({\tilde x}b|{\tilde a}p)$\\\hline
  3 & \=4 & N & Y &
  $(-1)^{n_k^\beta+n_K^\alpha+n_K^\beta+1}\sum_{ab}<i_\alpha|\hat a^+_{a\alpha}|k_\alpha>$\\
  &&&&$<i_\beta|\hat a^+_{b\beta}|k_\beta>\sum_{px}f_p(IK)f_x(JL)({\tilde b}x|{\tilde a}p)$\\\hline
  3 & 5 & Y  &  N & $\delta_{i_\beta
  k_\beta}(-1)^{n_K^\alpha+n_K^\beta+1}\sum_{c}<i_\alpha|\hat
  a_{c\alpha}|k_\alpha>\sum_pf_p(IK)g_{cp}(LJ)$\\ \hline
  3 & 7 & Y  &  N & $\delta_{i_\alpha
  k_\alpha}(-1)^{n_k^\alpha+n_K^\alpha+n_K^\beta}\sum_{c}<i_\beta|\hat
  a_{c\beta}|k_\beta>\sum_pf_p(IK)g_{pc}(LJ)$\\ \hline
  \end{tabular}
  \end{table}

  \begin{table}[h]
  \centerline{{ Table 3} continued}\vspace{10pt}
  \begin{tabular}{|c|c|c|c|c|}
  \hline
  Old case ($IK$) & New case ($IiKk$) & $(Ii)\le(Kk)$? & $J\le L$?
  & $<IiJ|\hat H|KkL>$ \\
  \hline\hline
  4 & 1 & YN & YN & $(-1)^{n_k^\alpha+n_K^\alpha+n_K^\beta+1}\left\{\delta_{JL}\left[
     \delta_{i_\alpha k_\alpha}\left(\sum_ag_a(IK)<i_\beta|\hat a^+_{a\beta}|k_\beta>+\right.\right.\right.$\\
  &&&& $\left.\sum_pf_p(IK)\sum_{a,b>c}<i_\beta|\hat
  a^+_{c\beta}\hat a^+_{b\beta}\hat
  a_{a\beta}|k_\beta>[({\tilde b}a|{\tilde c}p)-({\tilde c}a|{\tilde b}p)]\right)+$\\
  &&&& $\left.\sum_pf_p(IK)\sum_{abc}<i_\beta|\hat
  a^+_{c\beta}|k_\beta><i_\alpha|\hat a^+_{b\alpha}\hat
  a_{a\alpha}|k_\alpha>({\tilde b}a|{\tilde c}p)\right]+$\\
  &&&& $\left.\delta_{i_\alpha
  k_\alpha}\sum_{cp}g_{cp}^\beta(JL)f_p(IK)<k_\beta|\hat
  a_{c\beta}|i_\beta>\right\}$\\
  \hline
  4 & 2 & Y  &  N & $\delta_{i_\beta
  k_\beta}(-1)^{n_K^\alpha+n_K^\beta+1}\sum_{c}<i_\alpha|\hat
  a^+_{c\alpha}|k_\alpha>\sum_pf_p(IK)g_{cp}(LJ)$\\ \hline
  4 & 3 & Y & N &
  $(-1)^{n_k^\beta+n_K^\alpha+n_K^\beta+1}\sum_{ab}<i_\beta|\hat a^+_{a\beta}|k_\beta>$\\
  &&&&$<i_\alpha|\hat a_{b\alpha}|k_\alpha>\sum_{px}f_p(IK)f_x(LJ)({\tilde x}b|{\tilde a}p)$\\\hline
  4 & \=3 & N & Y &
  $(-1)^{n_k^\beta+n_K^\alpha+n_K^\beta}<i_\beta|\hat a^+_{a\beta}|k_\beta>$\\
  &&&&$<i_\alpha|\hat a^+_{b\alpha}|k_\alpha>\sum_{px}f_p(IK)f_x(JL)({\tilde b}x|{\tilde a}p)$\\\hline
  4 & 4 & Y & N &
  $(-1)^{n_k^\beta+n_k^\alpha+n_K^\alpha+n_K^\beta+1}\left\{\right.$\\
  &&&& $\delta_{ik}\left(\sum_pf_p(IK)f_p(LJ)+\sum_xg_x(IK)f_x(LJ)\right)+$\\
  &&&& $\delta_{i_\beta k_\beta}\sum_{ab}<i_\alpha|\hat
  a_{a\alpha}^+\hat
  a_{b\alpha}|k_\alpha>\sum_{px}f_p(IK)f_x(LJ)({\tilde x}p|{\tilde a}b)+$\\
  &&&& $\left.\delta_{i_\alpha k_\alpha}\sum_{ab}<i_\beta|\hatÜ
  a_{a\beta}^+\hat
  a_{b\beta}|k_\beta>\sum_{px}f_p(IK)f_x(LJ)[({\tilde x}p|{\tilde a}b)-({\tilde a}p|{\tilde x}b)]\right\}$\\\hline
  4 & \=4 & N & Y &
  $(-1)^{n_k^\beta+n_k^\alpha+n_K^\alpha+n_K^\beta}\delta_{i_\alpha k_\alpha}$\\
  &&&&$\sum_{a>b}<i_\beta|\hat a^+_{b\beta}\hat
  a^+_{a\beta}|k_\beta>\sum_{px}f_p(IK)f_x(JL)[({\tilde a}p|{\tilde b}x)-({\tilde b}p|{\tilde a}x)]$\\\hline
  4 & 6 & Y  &  N & $\delta_{i_\alpha
  k_\alpha}(-1)^{n_K^\alpha+n_K^\beta+n_k^\alpha+1}\sum_{c}<i_\beta|\hat
  a_{c\beta}|k_\beta>\sum_pf_p(IK)g_{cp}(LJ)$\\ \hline
  4 & 7 & Y  &  N & $\delta_{i_\beta
  k_\beta}(-1)^{n_K^\alpha+n_K^\beta+1}\sum_{c}<i_\alpha|\hat
  a_{c\alpha}|k_\alpha>\sum_pf_p(IK)g_{cp}(LJ)$\\ \hline\hline
  5 & 1 & YN & YN & $\delta_{JL}\delta_{i_\beta k_\beta}\sum_{a>b}
                    <i_\alpha|\hat a^+_{b\alpha}\hat a^+_{a\alpha}|k_\alpha>g_{ab}(IK)$\\ \hline
  5 & 3 & Y  &  N & $\delta_{i_\beta
  k_\beta}(-1)^{n_k^\alpha+n_k^\beta+1}\sum_{c}<i_\alpha|\hat
  a^+_{c\alpha}|k_\alpha>\sum_xf_x(LJ)g_{cx}(IK)$\\ \hline
  5 & 5 & Y  &  N & $\delta_{ik}\sum_{x>y}g_{xy}(IK)f_{xy}(LJ)$ \\
  \hline\hline
  6 & 1 & YN & YN & $\delta_{JL}\delta_{i_\alpha k_\alpha}
                    \sum_{a>b}<i_\beta|\hat a^+_{b\beta}\hat a^+_{a\beta}|k_\beta>g_{ab}(IK)$\\ \hline
  6 & 4 & Y  &  N & $\delta_{i_\alpha
  k_\alpha}(-1)^{n_k^\beta+1}\sum_{c}<i_\beta|\hat
  a^+_{c\beta}|k_\beta>\sum_xf_x(LJ)g_{cx}(IK)$\\ \hline
  6 & 6 & Y  &  N & $\delta_{ik}\sum_{x>y}g_{xy}(IK)f_{xy}(LJ)$ \\
  \hline\hline
  7 & 1 & YN & YN & $\delta_{JL}(-1)^{n_k^\alpha+1}\sum_{ab}<i_\alpha|\hat a^+_{a\alpha}|k_\alpha>
                    <i_\beta|\hat a^+_{b\beta}|k_\beta>g_{ab}(IK)$\\ \hline
  7 & 3 & Y  &  N & $\delta_{i_\alpha
  k_\alpha}(-1)^{n_k^\beta}\sum_{c}<i_\beta|\hat
  a^+_{c\beta}|k_\beta>\sum_xf_x(LJ)g_{xc}(IK)$\\ \hline
  7 & 4 & Y  &  N & $\delta_{i_\beta
  k_\beta}(-1)^{n_k^\alpha+n_k^\beta+1}\sum_{c}<i_\alpha|\hat
  a^+_{c\alpha}|k_\alpha>\sum_xf_x(LJ)g_{cx}(IK)$\\ \hline
  7 & 7 & Y  &  N & $\delta_{ik}\sum_{xy}g_{xy}(IK)f_{xy}(LJ)$ \\
  \hline
 \end{tabular}
 \end{table}

\section{Extended `primitive' matrix elements}\label{app3}
Here we give the working expressions for `primitive' matrix
elements in the extended basis $Ii,Kk$. The notations are the same
as in Appendix \ref{app2}. As cited in the text, these matrix
elements are not stored. Instead they are calculated `on the fly'
and then contracted with vectors coefficients produced by SVD
procedure to give final matrix elements for the reduced $M$ state
space. The full list of nonzero matrix elements is given in the
Table \ref{tab4}.

\begin{table}[h]
  \centering
  \caption{Extended `primitive' matrix elements for DMRG}\label{tab4}
\begin{tabular}{|c|c|c|c|}
  \hline
  Old case ($IK$) & New case ($IiKk$) & $(Ii)\le(Kk)$? & New MEs \\
  \hline\hline
  1 & 1 & YN & $\begin{array}{c} H_{IiKk}^{Aa}=\delta_{ik}H_{IK}^A+\delta_{IK}H_{ik}^a+\\ \sum_{ab}\big[
  \delta_{i_\beta k_\beta}g_{ab}^\alpha(IK)<i_\alpha|\hat a^+_{a\alpha}\hat a_{b\alpha}|k_\alpha>
  +\\ \delta_{i_\alpha k_\alpha}g_{ab}^\beta(IK)<i_\beta|\hat a^+_{a\beta}\hat a_{b\beta}|k_\beta>\big]\\
  f_{pq}^{\alpha,\beta}(IiKk)=\delta_{ik}f_{pq}^{\alpha,\beta}(IK)\\
  f_{ab}^{\alpha}(IiKk)=\delta_{IK}\delta_{i_\beta k_\beta}<i_\alpha|\hat a^+_{a\alpha}\hat a_{b\alpha}|k_\alpha>\\
  f_{ab}^{\beta}(IiKk)=\delta_{IK}\delta_{i_\alpha k_\alpha}<i_\beta|\hat a^+_{a\beta}\hat a_{b\beta}|k_\beta> \end{array}$ \\
  \hline
  1 & 2 & Y & $f_{ab}(IiKk)=\delta_{IK}(-1)^{n_k^\alpha}<i_\alpha|\hat a^+_{a\alpha}|k_\alpha><i_\beta|\hat a_{b\beta}|k_\beta>$ \\
  \hline
  1 & 3 & Y &  $\begin{array}{c} f_a(IiKk)=\delta_{IK}(-1)^{n_K^\alpha+n_K^\beta}\delta_{i_\beta k_\beta}
  <i_\alpha|\hat a_{a\alpha}|k_\alpha>\\ f_x(IiKk)=(-1)^{n_K^\alpha+n_K^\beta}\big[ \delta_{IK}f_x(ik)
  +\sum_ag_{xa}^\alpha(IK)\delta_{i_\beta k_\beta}f_a(i_\alpha k_\alpha)\big]\\ \end{array}$\\
  \hline
  1 & 4 & Y &  $\begin{array}{c} f_a(IiKk)=\delta_{IK}(-1)^{n_K^\alpha+n_K^\beta+n_k^\alpha}\delta_{i_\alpha k_\alpha}
  <i_\beta|\hat a_{a\beta}|k_\beta>\\ f_x(IiKk)=(-1)^{n_K^\alpha+n_K^\beta}\big[ \delta_{IK}f_x(ik)
  +(-1)^{n_k^\alpha}\sum_ag_{xa}^\beta(IK)\delta_{i_\alpha k_\alpha}f_a(i_\beta k_\beta)\big]\\ \end{array}$\\
  \hline
  1 & 5 & Y &  $f_{ab}(IiKk)=\delta_{IK}\delta_{i_\beta k_\beta}f_{ab}(i_\alpha k_\alpha)$\\
  \hline
  1 & 6 & Y &  $f_{ab}(IiKk)=\delta_{IK}\delta_{i_\alpha k_\alpha}f_{ab}(i_\beta k_\beta)$\\
  \hline
  1 & 7 & Y &  $f_{ab}(IiKk)=(-1)^{n_k^\alpha}\delta_{IK}<i_\alpha|\hat a_{a\alpha}|k_\alpha><i_\beta|\hat a_{b\beta}|k_\beta>$\\
  \hline
  \hline
  2 & 1 & YN & $H_{IiKk}^{Aa}=(-1)^{n_k^\alpha+1}\sum_{ab}g_{ab}(IK)<i_\alpha|\hat a_{a\alpha}|k_\alpha><i_\beta|\hat a^+_{b\beta}|k_\beta>$ \\
  \hline
  2 & 2 & Y &  $f_{pq}(IiKk)=\delta_{ik}f_{pq}(IK)$\\
  \hline
  2 & \=3 & N &  $f_x(KkIi)=\delta_{i_\alpha k_\alpha}(-1)^{n_K^\alpha+n_K^\beta+n_k^\alpha}\sum_{c}g_{xc}(IK)<i_\beta|\hat a^+_{c\beta}|k_\beta>$\\
  \hline
  2 & 4 & Y &  $f_x(IiKk)=\delta_{i_\beta k_\beta}(-1)^{n_K^\alpha+n_K^\beta+1}\sum_{c}g_{cx}(IK)<i_\alpha|\hat a_{c\alpha}|k_\alpha>$\\
  \hline
  \hline
  3 & 1 & YN &  $\begin{array}{c}H_{IiKk}^{Aa}=(-1)^{n_K^\alpha+n_K^\beta+1}\big[
  \sum_pf_p(IK)f_p(ki)+\sum_ag_a(IK)f_a(ki)\big]\\ f_{cp}(IiKk)^\alpha=(-1)^{n_K^\alpha+n_K^\beta+1}\delta_{i_\beta k_\beta}
  <i_\alpha|\hat a^+_{c\alpha}|k_\alpha>f_p(IK) \end{array} $ \\
  \hline
  3 & \=2 & N &  $f_{pc}(KkIi)=(-1)^{n_K^\alpha+n_K^\beta+n_k^\alpha+1}\delta_{i_\alpha k_\alpha}<i_\beta|\hat a^+_{c\beta}|k_\beta>f_p(IK)$\\
  \hline
  3 & 3 & Y &  $\begin{array}{c} f_p(IiKk)=\delta_{ik}f_p(IK)\\
  f_x(IiKk)=\delta_{ik}f_x(IK)+\\
  \delta_{i_\alpha k_\alpha}\sum_{ab}<i_\beta|\hat a^+_{a\beta}\hat a_{b\beta}|k_\beta>
  \sum_pf_p(IK)({\tilde x}p|{\tilde a}b)+\\
  \delta_{i_\beta k_\beta}\sum_{ab}<i_\alpha|\hat a^+_{a\alpha}\hat a_{b\alpha}|k_\alpha>
  \sum_pf_p(IK)\big[({\tilde x}p|{\tilde a}b)-({\tilde a}p|{\tilde x}b)\big]\end{array}$\\
  \hline
  3 & \=3 & N & $f_x(KkIi)=\delta_{i_\beta k_\beta}\sum_{a>b}<i_\alpha|\hat a^+_{b\alpha}\hat a^+_{a\alpha}|k_\alpha>
  \sum_pf_p(IK)\big[({\tilde a}p|{\tilde b}x)-({\tilde a}x|{\tilde b}p)\big]$ \\
  \hline
  3 & 4 & Y & $f_x(IiKk)=(-1)^{n_k^\alpha+1}\sum_{ab}<i_\alpha|\hat a^+_{a\alpha}|k_\alpha>
  <i_\beta|\hat a_{b\beta}|k_\beta>
  \sum_pf_p(IK)({\tilde x}b|{\tilde a}p)$ \\
  \hline
  3 & \=4 & N & $f_x(KkIi)=(-1)^{n_k^\alpha+1}\sum_{ab}<i_\alpha|\hat a^+_{a\alpha}|k_\alpha>
  <i_\beta|\hat a^+_{b\beta}|k_\beta>
  \sum_pf_p(IK)({\tilde b}x|{\tilde a}p)$ \\
  \hline
  3 & 5 & Y &  $f_{cp}(IiKk)=(-1)^{n_K^\alpha+n_K^\beta+1}\delta_{i_\beta k_\beta}<i_\alpha|\hat a_{c\alpha}|k_\alpha>f_p(IK)$\\
  \hline
  3 & 7 & Y &  $f_{pc}(IiKk)=(-1)^{n_K^\alpha+n_K^\beta+n_k^\alpha}\delta_{i_\alpha k_\alpha}<i_\beta|\hat a_{c\beta}|k_\beta>f_p(IK)$\\
  \hline
  \hline
  4 & 1 & YN &
  $\begin{array}{c}H_{IiKk}^{Aa}=(-1)^{n_K^\alpha+n_K^\beta+1}\big[
  \sum_pf_p(IK)f_p(ki)+\sum_ag_a(IK)f_a(ki)\big]\\ f_{cp}(IiKk)^\beta=(-1)^{n_K^\alpha+n_K^\beta+n_k^\alpha+1}\delta_{i_\alpha k_\alpha}
  <i_\beta|\hat a^+_{c\beta}|k_\beta>f_p(IK) \end{array} $ \\
  \hline
  4 & 2 & Y &  $f_{cp}(IiKk)=(-1)^{n_K^\alpha+n_K^\beta+1}\delta_{i_\beta k_\beta}<i_\alpha|\hat a^+_{c\alpha}|k_\alpha>f_p(IK)$\\
  \hline
  4 & 3 & Y & $f_x(IiKk)=(-1)^{n_k^\alpha}\sum_{ab}<i_\alpha|\hat a_{b\alpha}|k_\alpha>
  <i_\beta|\hat a^+_{a\beta}|k_\beta>
  \sum_pf_p(IK)({\tilde x}b|{\tilde a}p)$ \\
  \hline
  4 & \=3 & N & $f_x(KkIi)=(-1)^{n_k^\alpha}\sum_{ab}<i_\alpha|\hat a^+_{b\alpha}|k_\alpha>
  <i_\beta|\hat a^+_{a\beta}|k_\beta>
  \sum_pf_p(IK)({\tilde b}x|{\tilde a}p)$ \\
  \hline
  4 & 4 & Y &  $\begin{array}{c} f_p(IiKk)=\delta_{ik}f_p(IK)\\
  f_x(IiKk)=\delta_{ik}f_x(IK)+\\
  \delta_{i_\beta k_\beta}\sum_{ab}<i_\alpha|\hat a^+_{a\alpha}\hat a_{b\alpha}|k_\alpha>
  \sum_pf_p(IK)({\tilde x}p|{\tilde a}b)+\\
  \delta_{i_\alpha k_\alpha}\sum_{ab}<i_\beta|\hat a^+_{a\beta}\hat a_{b\beta}|k_\beta>
  \sum_pf_p(IK)\big[({\tilde x}p|{\tilde a}b)-({\tilde a}p|{\tilde x}b)\big]\end{array}$\\
  \hline
  4 & \=4 & N & $f_x(KkIi)=\delta_{i_\alpha k_\alpha}\sum_{a>b}<i_\beta|\hat a^+_{b\beta}\hat a^+_{a\beta}|k_\beta>
  \sum_pf_p(IK)\big[({\tilde a}p|{\tilde b}x)-({\tilde a}x|{\tilde b}p)\big]$ \\
  \hline
  4 & 6 & Y &  $f_{cp}(IiKk)=(-1)^{n_K^\alpha+n_K^\beta+n_k^\alpha+1}\delta_{i_\alpha k_\alpha}<i_\beta|\hat a_{c\beta}|k_\beta>f_p(IK)$\\
  \hline
  4 & 7 & Y &  $f_{cp}(IiKk)=(-1)^{n_K^\alpha+n_K^\beta+1}\delta_{i_\beta k_\beta}<i_\alpha|\hat a_{c\alpha}|k_\alpha>f_p(IK)$\\
  \hline
  \hline
\end{tabular}\end{table}
\begin{table}
  \centering
  Table 4 continued
\begin{tabular}{|c|c|c|c|}
  \hline
  Old case ($IK$) & New case ($IiKk$) & $(Ii)\le(Kk)$? & New MEs \\
  \hline\hline
  5 & 1 & YN & $H_{IiKk}^{Aa}=\sum_{a>b}g_{ab}(IK)f_{ab}(ki)$ \\
  \hline
  5 & 3 & Y  &
  $f_x(IiKk)=(-1)^{n_K^\alpha+n_K^\beta}\delta_{i_\beta
  k_\beta}\sum_{c}<i_\alpha|\hat a^+_{c\alpha}|k_\alpha>g_{cx}(IK)$\\
  \hline
  5 & 5 & Y  & $f_{pq}(IiKk)=\delta_{ik}f_{pq}(IK)$\\
  \hline\hline
  6 & 1 & YN & $H_{IiKk}^{Aa}=\sum_{a>b}g_{ab}(IK)f_{ab}(ki)$ \\
  \hline
  6 & 4 & Y  &
  $f_x(IiKk)=(-1)^{n_K^\alpha+n_K^\beta+n_k^\alpha}\delta_{i_\alpha
  k_\alpha}<i_\beta|\hat a^+_{c\beta}|k_\beta>g_{cx}(IK)$\\
  \hline
  6 & 6 & Y  & $f_{pq}(IiKk)=\delta_{ik}f_{pq}(IK)$\\
  \hline\hline
  7 & 1 & YN & $H_{IiKk}^{Aa}=\sum_{ab}g_{ab}(IK)f_{ab}(ki)$ \\
  \hline
  7 & 3 & Y  &
  $f_x(IiKk)=(-1)^{n_K^\alpha+n_K^\beta+n_k^\alpha+1}\delta_{i_\alpha
  k_\alpha}\sum_{c}<i_\beta|\hat a^+_{c\beta}|k_\beta>g_{xc}(IK)$\\
  \hline
  7 & 4 & Y  &
  $f_x(IiKk)=(-1)^{n_K^\alpha+n_K^\beta}\delta_{i_\beta
  k_\beta}\sum_{c}<i_\alpha|\hat a^+_{c\alpha}|k_\alpha>g_{cx}(IK)$\\
  \hline
  7 & 7 & Y  & $f_{pq}(IiKk)=\delta_{ik}f_{pq}(IK)$\\
  \hline\hline
\end{tabular}\end{table}

\section*{Acknowledgements}
All authors are grateful to the University of Bologna (Pluriennale
2002), to the MIUR of Italy and to the EEC (HPRN-CT-1999-00005)
for financial support; A. O. M. in particular for a fellowship
from the University of Bologna.


\newpage
\begin{thebibliography}{99}
\normalsize
\bibitem{White1} S. R. White, Phys. Rev. Lett. \textbf{69}, 2863
(1992).
\bibitem{White2} S. R. White and R. M. Noack, Phys. Rev. Lett.
\textbf{68}, 3487 (1992).
\bibitem{WhiteDMRG93} S. R. White, Phys. Rev. B \textbf{48}, 10345 (1993).
\bibitem{PariserFano} G. Fano, F. Ortolani and L. Ziosi, J. Chem.
Phys. \textbf{108}, 9246 (1998).
\bibitem{Bendaz} G. L. Bendazzoli, S. Evangelisti, G. Fano, F. Ortolani
and L. Ziosi,
  J. Chem. Phys. \textbf{110}, 1277 (1999).
\bibitem{WhiteDMRGqc1} S. R. White and R. L. Martin, J. Chem. Phys.
\textbf{110}, 4127 (1999).
\bibitem{WhiteDMRGqc2} S. Daul, I. Ciofini, C. Daul and S. R. White,
Int. J. Quantum Chem. \textbf{79}, 331 (2000).
 \bibitem{FanoLing} A. O. Mitrushenkov, G. Fano, F. Ortolani, R.
Linguerri and P. Palmieri, J. Chem. Phys. \textbf{115}, 6815
(2001).
 \bibitem{Gordon} Garnet Kin-Lic Chan and Martin Head-Gordon, J. Chem. Phys.
\textbf{116}, 4462 (2002).
\bibitem{Hess} Ors Legeza,Johannes Roder and Bernd A.Hess, to be published
in Mol. Phys.

\bibitem{Thorst} T.Thorsteinsson and D.L.Cooper, An Overview of the CASVB
Approach to Modern Valence Bond Calculations, in ``Quantum Systems in
Chemistry and Physics. Vol. 1: Basic problems and models systems'', ed.
A.Hernandez-Laguna,J.Maruani, R.McWeeny and S.Wilson (Kluwer,Dordrecht);
303 (2000).


\bibitem{blas1}C.
L. Lawson, R. J. Hanson, D. Kincaid, and F. T. Krogh, Basic Linear
Algebra Subprograms for FORTRAN usage, ACM Trans. Math. Soft., 5
(1979), pp. 308--323.
\bibitem{blas2}J. J. Dongarra, J. Du Croz, S. Hammarling, and R. J. Hanson, An
extended set of FORTRAN Basic Linear Algebra Subprograms, ACM
Trans. Math. Soft., 14 (1988), pp. 1--17.
\bibitem{blas3}J. J. Dongarra, J. Du Croz, S. Hammarling, and R. J. Hanson,
Algorithm 656: An extended set of FORTRAN Basic Linear Algebra
Subprograms, ACM Trans. Math. Soft., 14 (1988), pp. 18--32.
\bibitem{blas4}J. J. Dongarra, J. Du Croz, I. S. Duff, and S. Hammarling, A set
of Level 3 Basic Linear Algebra Subprograms, ACM Trans. Math.
Soft., 16 (1990), pp. 1--17.
\bibitem{blas5}J. J. Dongarra, J. Du Croz, I. S. Duff, and S. Hammarling,
Algorithm 679: A set of Level 3 Basic Linear Algebra Subprograms,
ACM Trans. Math. Soft., 16 (1990), pp. 18--28.

\bibitem{Douall} McDouall,J.J.W.,Theoret.chim.Acta \textbf{83}, 339 (1992);
\textbf{85}, 395 (1993).
\bibitem{Thorstein} T.Thorsteinsson and D.L.Cooper, Mol. Phys. \textbf{93},
 663 (1998).
 \bibitem{Vorst}Henk A. van der Vorst, Computational methods for Large
Eigenvalue Problems,
http://www.math.ruu.nl/people/vorst/lecture.html
\end{thebibliography}
\end{document}